\documentclass[final]{siamltex}
\usepackage{amssymb,amsmath}
\usepackage{graphicx}
\title{Descriptor approach for eliminating spurious eigenvalues in hydrodynamic equations}

\author{M. Lisa Manning\thanks{Center for Theoretical Science, Princeton University, NJ 08544({\tt lm2@princeton.edu}).}
        \and B. Bamieh \thanks{Department of Mechanical Engineering, University of California, Santa Barbara, CA 93106}
	\and J. M. Carlson\thanks{Department of Physics, University of California, Santa Barbara, CA 93106-9530}}

\begin{document}

\maketitle

\begin{abstract}
We describe a general framework for avoiding spurious eigenvalues --- unphysical unstable eigenvalues that often occur in hydrodynamic stability problems. In two example problems, we show that when system stability is analyzed numerically using {\em descriptor} notation, spurious eigenvalues are eliminated. Descriptor notation is a generalized eigenvalue formulation for differential-algebraic equations that explicitly retains algebraic constraints.  We propose that spurious eigenvalues are likely to occur when algebraic constraints are used to analytically reduce the number of independent variables in a differential-algebraic system of equations before the system is approximated numerically. In contrast, the simple and easily generalizable descriptor framework simultaneously solves the differential equations and algebraic constraints and is well-suited to stability analysis in these systems.
\end{abstract}

\begin{keywords}
spurious eigenvalue, descriptor, differential algebraic, spectral method, incompressible flow, hydrodynamic stability, generalized eigenvalue, collocation
\end{keywords}


\pagestyle{myheadings}
\thispagestyle{plain}
\markboth{M. L. MANNING, B. BAMIEH, AND J. M. CARLSON}{DESCRIPTOR APPROACH FOR ELIMINATING SPURIOUS EIGENVALUES \ldots}

\section{Introduction}

 Spurious eigenvalues are unphysical, numerically-computed eigenvalues with large positive real parts that often occur in hydrodynamic stability problems. We propose that these unphysical eigenvalues occur when the incompressible Navier Stokes equations are analytically reduced -- i.e., the algebraic constraints are used to reduce the number of independent variables before the system is approximated using spectral methods. 

An alternative approach to analyzing differential-algebraic equations is the {\em descriptor} framework, posed as a generalized eigenvalue problem, which explicitly retains the algebraic constraints during the numerical computation of eigenvalues.  We reformulate two common hydrodynamic stability problems using descriptor notation and show that this method of computation avoids the spurious eigenvalues generated by other methods.  The descriptor formulation is a simple, robust framework for eliminating spurious eigenvalues that occur in hydrodynamic stability analysis.  Additionally, this formulation reduces the order of the numerically approximated differential operators and accommodates complex boundary conditions(BCs), such as a fluid interacting with a flexible wall.

   Resolving the spectrum of hydrodynamic operators is critical for time integration, linear stability~\cite{Drazin} and transient growth analyses~\cite{SchmidHen,Butler,Trefethen,Jovanovic,Gustavsson2,Farrell,KimLim,Schmid,Hogberg} of fluid flows. Spurious eigenvalues often arise in these spectral numerical computations describing incompressible fluids and must be identified or eliminated. Although the term ``spurious'' is applied to many numerical anomalies, it is used here to refer exclusively to large, positive eigenvalues that arise due to application of extra boundary conditions. These are distinct, for example, from spurious pressure modes that occur when the momentum equation is collocated at interior Chebyshev-Gauss-Lobatto nodes~\cite{Phillips}.

 Researchers have developed special methods to avoid or filter these modes and uncover the true spectrum of the model problem.  Perhaps the first description of these unphysical values is given by Gottlieb and Orszag~\cite{Gottlieb}.  Many other researchers have encountered similar modes~\cite{Brenier,Straughan,Zebib1} and developed methods for avoiding~\cite{Gottlieb,Huang,Gardner,McFadden,Zebib2} or filtering~\cite{Orszag,Bewley} them.  These methods for avoiding spurious modes are specific to very special {\em clamped} BCs where homogeneous Dirichlet and Neumann conditions hold at the boundaries. We emphasize that the descriptor method can be implemented using widely available numerical libraries and is easily generalized to non-clamped BCs.

  This paper is divided as follows:   Section~\ref{descriptor} discusses the general descriptor framework and Sections~\ref{orr1}~and~\ref{1D} develop this method for specific examples, the Orr-Sommerfeld operator and Gottlieb and Orszag's one-dimensional potential flow model~\cite{Gottlieb}.   Section~\ref{discussion} discusses how infinite eigenvalues arise in formulations of the incompressible fluid equations, and explains how the descriptor formulation explicitly accounts for these eigenvalues. We also discuss why methods that include analytical transformations might generate spurious eigenvalues, as well as benefits and drawbacks of descriptor formulations.

\section{The descriptor framework}
\label{descriptor}

Descriptor notation can be used to describe any dynamical system where a set of differential-algebraic equations is reduced to a set of differential equations only. In this paper we focus on two models for incompressible fluid flow: the 2D Navier-Stokes equations which generate the Orr-Sommerfeld (OS) operator, and Gottlieb and Orszag's 1D potential flow model.

 Descriptor notation was developed in the control theory community~\cite{Luenberger,Sincovec,Kunkel} to describe and analyze systems of differential-algebraic equations. In descriptor form, the differential time operator is preceded by a square, possibly singular matrix: 
\begin{equation}
E \frac{\partial}{\partial t} \phi = A \phi .
\end{equation}
In descriptor systems, stability is determined by the generalized eigenvalues of the ($A,E$) system, which is the ratio of the pair $( \alpha, \beta)$ where $\beta A \boldsymbol{u} = \alpha E \boldsymbol{u}$ for some non-zero vector $\boldsymbol{u}$.   While traditional eigenvalues are never infinite, descriptor system can have (many) infinite eigenvalues.  If $E$ contains a zero row corresponding to an algebraic constraint, there will be an infinite eigenvalue corresponding to the infinitely fast dynamics of that constraint. 

Let us assume that we have a system of differential-algebraic equations for $n$ fields. Let there be $m$ algebraic constraints, and $k = n-m$ equations that contain a differential time operator. Physical systems are often modeled by differential-algebraic equations of this form because algebraic constraints often arise as approximations to differential equations for quickly equilibrating variables.

  Let the vector $\boldsymbol{v}$ contain the $k$ fields that are acted upon by the differential time operator, and $p$ contain the remaining $m$ fields. Each of the $n$ fields can be discretized using $N$ points for each field, which results in a finite-dimensional system of differential algebraic equations that can be written as follows:

\begin{eqnarray}
\begin{bmatrix} I & 0 \\ 0 & 0 \end{bmatrix} \begin{bmatrix}\dot{\boldsymbol v} \\ \dot{p} \end{bmatrix} &=& \begin{bmatrix} A_{11} & A_{12} \\ A_{21} & A_{22} \end{bmatrix} \begin{bmatrix} {\boldsymbol v} \\ p \end{bmatrix} ;\\
E \dot{\phi} &\equiv& A \phi .
\end{eqnarray}

  The discretized operators $A$ and $E$ are square matrices with $(n \times N)^{2}$ entries. If the original equations are partial differential equations, the operator $A$ contains spatial derivatives that require BCs.  Any BC can be incorporated into the numerical solution using boundary bordering in a straightforward manner~\cite{Boyd_book}.  For example, if $A$ contains spatial derivatives up to order $G$ for each of the $k$ fields in $\boldsymbol{v}$, then $G$ BCs are required for each field and are enforced by using ($G \times k$) rows of $A$ as additional algebraic constraints.  We then solve for the generalized eigenvalues of the ($A,E$) pair. We assume the pair is {\em regular}, which means that det$(sE -A)$ is not identically zero for all~$s$.   There are many numerical routines that solve for the generalized eigenvalues of regular matrix pairs when one of the matrices is singular (see references in~\cite{Watkins}and~\cite{Kagstrom}). A widely available routine is the MATLAB (LAPACK) '{\bf QZ}' algorithm~\cite{Moler}. Note that $E$ contains ($m N$) zero rows corresponding to the original algebraic constraints and ($k G$) zero rows corresponding to the boundary bordering constraints, resulting in an ($A,E$) pair with $(m N +  k G)$ infinite generalized eigenvalues and $ (k N - k G)$ finite eigenvalues. Although the algebraic constraints are approximated using boundary bordering, descriptor notation ensures that the generalized eigenvalues corresponding to these constraints are not approximated -- they are formally infinite.

Our method is in contrast to other methods where the algebraic constraints are removed analytically. In these methods, the system is  analytically converted to a system of equations that contains only fields in $v$, and consequently, a subset of the algebraic constraints on $p$ are converted to additional BCs on $v$. Discretization and enforcement of the new BCs may lead to an approximate enforcement of the original algebraic constraints, resulting in spurious eigenvalues.

 One example of this over-specification of the BCs occurs in the derivation of the OS operator. Assume that $A_{12}$ contains a first order spatial derivative and $A_{22}$ is zero. We eliminate $p$, using a method described for the OS operator in Appendix~\ref{orr_som_app}~(Eqs.~\ref{RigidOrig}-\ref{OrrEnd}), and the resulting system of equations contains a spatial derivative of order $2 + G$ acting on the fields in $\boldsymbol{v}$. Because the derivative operator is two orders higher than before, the system requires two new BCs. These extra BCs are determined by numerically approximating the algebraic constraints at the boundary. However, the algebraic constraints were used to eliminate $p$ and were already evaluated at the boundary in the analytical computation. In several spectral methods, the algebraic constraints at the boundary are enforced twice, with slightly different numerical approximations in each case.

 We now illustrate how the descriptor method avoids this over-specification in two example problems.

\section{Example: Orr-Sommerfeld Operator}
\label{orr1}  
  The Navier-Stokes equations for a compressible, viscous fluid can be written~\cite{Lifshitz}:
\begin{eqnarray}
\label{mom_cons}
 \frac{\partial {\boldsymbol v}}{\partial_t} &=& - \frac{1}{\rho} \nabla p - \left({\boldsymbol v} \cdot \nabla \right) {\boldsymbol v} + \nu \Delta {\boldsymbol v} + \left(\xi + \frac{1}{3} \nu \right) \nabla ( \nabla \cdot{\boldsymbol v} ) ; \\
\label{mass_cons}
 0 &=& \frac{\partial \rho}{\partial t} + \nabla \cdot \left (\rho {\boldsymbol v}\right),
\end{eqnarray}
where ${\boldsymbol v}$ represents the three components of the velocity, $p$ is the fluid pressure, $\rho$ is the the mass density, and $\nu$ and $\xi$ are the first and second kinematic viscosities, respectively. An additional thermodynamic constraint such as constant entropy relates the variables ${\boldsymbol v}, \rho$ and $p$ and closes the system of equations.

 As the system approaches the incompressible limit, the partial differential equation expressing conservation of mass (Eq.~\ref{mass_cons}) becomes increasingly stiff, and in the limit a differential equation in time is replaced by an algebraic constraint, 
\begin{equation}
\label{div_const}
\nabla \cdot {\boldsymbol v} = 0.
\end{equation}
 As the equation for the mass density~\ref{mass_cons} becomes more stiff the eigenvalues of the operator will have larger real parts, and in the limit the eigenvalues will be infinite.  In this case the pressure can be thought of as a Lagrange multiplier that instantaneously satisfies the divergence constraint and no thermodynamic equation is required to relate the pressure and density. Therefore Eqs.~(\ref{mom_cons},~\ref{div_const}) alone are the Navier-Stokes equations for an incompressible fluid.

   Numerical solutions to the linearized, incompressible equations of motion have traditionally been determined by combining the Laplacian and divergence of the NS equations.   This allows the pressure to be rewritten in terms of the fluid velocities, thereby reducing the set of four differential algebraic equations to two differential equations. For later reference, these equations are derived using operator multiplication in Appendix~\ref{orr_som_app}. In planar channel flow this results in the OS equation for the wall-normal velocity and the Squire equation for the wall-normal vorticity~\cite{Kim,Gustavsson}. Descriptor notation does not require this reduction, and does not generate the fourth order OS operator.

   In most two-dimensional flow models the system is assumed to be translation invariant in the spanwise($z$) direction, so the coupling between the vorticity and velocity is zero and eigenvalues of the OS operator determine the system stability. In order to compare our results with previous studies, we focus here on the stability of the OS operator alone. 

 The OS operator contains the term $\Delta^{2}$, which is a fourth-order spatial derivative in the wall-normal direction and requires four BCs. Two BCs are simply the no-slip conditions from the original equations, $v_y(\pm 1) =0$. The remaining two BCs arise from the divergence constraint:
\begin{equation}
\label{Neumann}
\left. \frac{\partial v_y}{\partial y} \right|_{y = \pm 1}  = -\left( \left. \frac{\partial v_x}{\partial x} \right|_{y = \pm 1}  + \left. \frac{\partial v_z}{\partial z}\right|_{y = \pm 1}  \right) = 0 + 0 ,
\end{equation}
where the last equality holds because $v_x$ and $v_z$ are constant (zero) in the streamwise ($x$)- and spanwise($z$)- directions at the boundary. Therefore, homogeneous Neumann BC on the wall-normal velocity is a direct consequence of the incompressible limit. This has important implications for numerical approximations, as well will discuss in Section~\ref{discussion}.  

Spurious eigenvalues arise frequently in the analysis of this OS operator~\cite{Gardner,Helgason,Zebib1}. While there are several methods for avoiding or filtering these eigenvalues, such as applying only Dirichlet conditions to the second order operator~\cite{Weideman,McFadden,Huang}, they are tailored to clamped boundary conditions and are not easily generalizable.  Our approach to solving the original system of Eqs.~(\ref{mom_cons},~\ref{div_const}) is more general. We avoid combining the two equations into a single system and instead write the system using descriptor notation. A similar analysis using descriptor notation was applied previously to incompressible Stokes flow in the context of systems control~\cite{Stykel1,Stykel2}, which we generalize to eliminate spurious eigenvalues. The NS equations~(\ref{mom_cons},~\ref{div_const}) can be written: 
\begin{eqnarray}
\label{des_Orr}
\begin{bmatrix} I & 0 \\ 0 & 0 \end{bmatrix} \begin{bmatrix}\dot{v} \\ \dot{p} \end{bmatrix} &=& \begin{bmatrix} A & Q \\ D & 0 \end{bmatrix} \begin{bmatrix} {\boldsymbol v} \\ p \end{bmatrix} ;\\
E \dot{\phi} &\equiv& \overline{A} \phi ;\\
\label{des_Orr3}
v_y(\pm 1) &=& v_x(\pm 1) \: = \:  v_z(\pm 1) = 0 .
\end{eqnarray}
The operators $A, Q$ and $D$ are defined in Appendix ~\ref{orr_som_app}.  

   In hydrodynamic channel flow, a Fourier transform can be taken in the translation invariant streamwise and spanwise directions.  The velocities have Dirichlet BCs at the channel boundaries, $y=\pm 1$, and we discretize the system of equations in the wall normal direction using Chebyshev collocation or other spectral methods. No-slip BCs are enforced for the velocities, but care must be taken to ensure that no BCs are applied to the pressure.

 For simplicity, boundary bordering is used to enforce Dirichlet BCs on each component of the velocity. We approximate each component of the velocity and the pressure by a vector of $N$ points. The BCs require that the first and last entry of each velocity vector is zero:
\begin{eqnarray}
v_{i}(y_0) &=& v_{i}(+1) =0 ;\\
v_{i}(y_{(N-1)}) &=& v_{i}(-1) =0 ;\\
i &=& {x, y, z}. \nonumber
\end{eqnarray}
This is equivalent to deleting the first and last columns of $\mbox{Cheb}^{(1)}$ and $\mbox{Cheb}^{(2)}$ that occur in the operators $D$ and $A$, respectively, in Eq.~\ref{des_Orr}. Additionally, time derivatives of the velocity evaluated at the boundary are zero $\dot{v}_{i}(y_0) =  \dot{v}_{i}(y_{N-1}) = 0$, which is equivalent to deleting the corresponding rows of $A$ and $Q$ in Eq.~\ref{des_Orr}.  As a result, each component of the velocity is represented by an $N-2$ column vector. Let $M = 3 \times (N-2)$. Then the operator $A$ is approximated by an $M$-by-$M$ matrix, $Q$ is a matrix that is $M$ rows by $N$ columns, $D$ is $N$ rows by $M$ columns. Note that there are no BCs imposed on $p$, which is still a vector of size $N$.  We then solve for the generalized eigenvalues of the ($\overline{A},E$) pair using the MATLAB (LAPACK) '{\bf QZ}' algorithm~\cite{Moler}. In the parameter range studied, the pair is {\em regular}, so that det$(sE -\overline{A})$ is not identically zero for all~$s$.  Because $E$ contains $N$ rows that are singular, there are $N$ infinite generalized eigenvalues and $M$ finite eigenvalues. 

Table~\ref{Orr_table}.1 lists the first three non-infinite eigenvalues calculated from the descriptor system given by Eqs.~\ref{des_Orr}-\ref{des_Orr3} with 34 collocation points, $\alpha=1$ and $R= 10\,000$.  These three eigenvalues match those calculated using other methods which have been identified as real, physical eigenmodes -- there are no spurious eigenvalues. For comparison, Table~\ref{Orr_table}.1 also lists the first three non-infinite eigenvalues generated using a traditional Orr-Sommerfeld Chebyshev Tau method. The traditional method generates a pair of large, positive, spurious eigenvalues.

\begin{table}
\begin{center}
\label{Orr_table}
\begin{tabular}{|l|l|l|l|}
\hline
{\bf Numerical Method} & $\boldsymbol{\lambda_1} $ & $\boldsymbol{\lambda_2}$ & $\boldsymbol{\lambda_3}$ \\ \hline
Descriptor Chebyshev Tau   & +.0037 & -0.0348 & -0.0350 \\ \hline
OS Chebyshev Tau & 97.557 & 85.735 & 0.037 \\ \hline
\end{tabular}
\caption{This table compares the first three non-infinite eigenvalues for the 2D NS equations with $\alpha=1$, $R=10\: 000$ for a descriptor vs. an Orr Sommerfeld(OS) spectral collocation scheme with $N$ = 34. The OS scheme generates two large, positive eigenvalues, while the descriptor scheme does not.}
\end{center}
\end{table}

A descriptor formulation can be used for any system of differential algebraic equations. As another example we now formulate Gottlieb and Orszag's one-dimensional potential flow model using descriptor notation and show that this method avoids spurious eigenvalues.

\section{Example: Incompressible limit of a 1D fluid model}
\label{1D}
 Spurious eigenvalues have been studied in greatest depth using the model problem of Gottlieb and Orszag~\cite{Gottlieb}. This is a model for a two-component, one-dimensional fluid flow at low Reynolds number, and is described by the following equations:
\begin{eqnarray}
\label{stream_system}
\frac{\partial \zeta}{\partial t} &=& \nu \frac{\partial^2 \zeta}{\partial x^2}; \\
\label{ss2}
\zeta &=& \frac{\partial^2 \psi}{\partial x^2}.
\end{eqnarray}
Here $\zeta$ is the vorticity and $\psi$ is the stream function defined by $(v_x, v_y) = (-\partial \psi / \partial y, \partial \psi / \partial x)$. The divergence constraint is automatically satisfied because  $ \nabla \cdot (v_x, v_y)~\equiv~0$ equates the mixed partial derivatives of $\psi$, which is always true for analytic $\psi$.  For a fluid between stationary rigid walls, no-slip conditions on the velocity at the boundary correspond to the following constraints on the stream function:
\begin{equation}
\psi(x=\pm1, t) = \psi_{x}(x = \pm 1, t) = 0. 
\end{equation}
The usual method used to find the eigenvalues of Eqs.~\ref{stream_system} combines the two equations into a single partial differential equation,
\begin{equation}
\psi_{xxt} = \nu \psi_{xxxx},
\end{equation}
then inverts the second order differential operator and represents the operators in terms of spectral differentiation matrices,
\begin{equation}
\psi_{t} = \nu (\partial_{xx})^{-1} \partial_{xxxx} \psi.
\end{equation}
The second order operator is rendered invertible through the application of two BCs. However, there are four BCs on $\psi$, so there is ambiguity as to which two should be applied. One choice uses basis recombination so that each of the basis functions individually satisfies all four BCs.  Dawkins, Dunbar and Douglass~\cite{Dawkins} have shown that a numerical scheme with Chebyshev basis functions generates large spurious eigenvalues, while a scheme with Legendre basis functions generate formally infinite eigenvalues.  They also show that Neumann BCs exactly match the form of the Legendre polynomials, and conclude that spurious eigenvalues are approximations to these infinite eigenvalues that occur when a Chebyshev basis is used instead of a Legendre basis.

 Because the BCs are algebraic constraints that correspond to infinitely fast pressure modes, the Legendre polynomials are an exact basis for approximating the algebraic constraints, and therefore exactly recover the infinite eigenvalues that correspond to these constraints. The Legendre basis recombination method is similar to the descriptor framework in that the infinite eigenvalues are exactly computed in each case.

  An alternate method for incorporating BCs is the traditional Chebyshev-Tau method, where boundary bordering is used to replace four terms in the Chebyshev expansion with four algebraic constraints.  These algebraic constraints are then used to reduce the total number of equations by four, and the eigenvalues of the resulting system of equations are computed~\cite{Gardner}.  This reduced set of differential equations has been shown to be equivalent to the Chebyshev basis recombination method described above~\cite{Dawkins}, and generates spurious eigenvalues, as shown in Table~\ref{compare_descriptor}.1. Various other approaches to solving the spurious eigenvalue problem involve imposing only the Dirichlet BCs on the second order differential operator~\cite{McFadden,Huang}. One such approach is Weideman and Reddy's~\cite{Weideman} spectral collocation scheme with modified clamped conditions, which is also listed in Table~\ref{compare_descriptor}.1.

In contrast to these approaches, the system Eqs.~(\ref{stream_system},~\ref{ss2}) can be written in descriptor form:
\begin{eqnarray}
\label{1dmat}
\begin{bmatrix} I & 0 \\ 0 & 0 \end{bmatrix} \begin{bmatrix} \dot{\zeta} \\ \dot{\psi} \end{bmatrix} &=& \begin{bmatrix} \nu \partial_{xx} & 0 \\ -I & \partial_{xx} \end{bmatrix} \begin{bmatrix} \zeta \\ \psi \end{bmatrix} ; \\
E  \begin{bmatrix} \dot{\zeta} \\ \dot{\psi} \end{bmatrix} &\equiv& A \begin{bmatrix} \zeta \\ \psi \end{bmatrix}.
\end{eqnarray} 
A second order operator acts on $\psi$ and there are four BCs for $\psi$, while a second order operator acts on $\zeta$ and there are no BCs on $\zeta$. This is a similar situation to the OS problem because the four conditions on $\psi$ come from a combination of the no-slip condition and the divergence constraint. 

Therefore we expect these four modes to have infinite eigenvalues and impose all the BCs on $\psi$ as algebraic constraints.  The resulting approximation to the bottom row of the matrix Eq.~\ref{1dmat} ($0 = -\zeta + \partial_{xx} \psi$) is:
\begin{eqnarray}
0 &=& \psi_0;  \\
\label{interior_des}
\boldsymbol{0} &=& -\delta_{ij} \zeta_{j} + \mbox{Cheb}^{(2)}_{ij} \quad \psi_j, j,i \in (1, N-2);\\
0 &=& \psi_{(N-1)},
\end{eqnarray}
and the approximation to the top row of Eq.~\ref{1dmat}, $\partial_t \zeta = \partial_{xx} \zeta$, is:
\begin{eqnarray}
0 &=& \mbox{Cheb}^{(1)}_{1j} \psi_j, \quad j \in (1, N-2); \\
\partial_t \zeta_i &=& \mbox{ Cheb}^{(2)}_{ij} \zeta_{j} , \quad i \in (1, N-2), j \in (0,N-1); \\
0 &=& \mbox{Cheb}^{(1)}_{(N-2)j} \psi_j  \quad j \in (1, N-2).
\end{eqnarray}
 With descriptor notation there are $(N-2)$ infinite eigenvalues corresponding to the $(N-2)$ interior points of the algebraic constraint Eq.~\ref{interior_des}.  The four new BCs correspond to four formally infinite eigenvalues in the numerical spectrum. This final descriptor system contains an $E$ matrix with $(N-2) + 4$ rows of zeros, and therefore the system has $N+2$ generalized infinite eigenvalues. If a Chebyshev-Tau method is used to discretize the differential operators, descriptor notation is equivalent to the method suggested by Gottlieb and Orszag to avoid spurious eigenvalues when they first posed this simple model~\cite{Gottlieb}, except that we use matrix methods for singular matrices instead of a shooting algorithm to determine eigenvalues. Again, the '{\bf QZ }' routine in MATLAB is used to compute the spectrum for this system of equations.  

Table~\ref{compare_descriptor}.1 compares the first three non-infinite numerical eigenvalues to the analytically computed eigenvalues for several different discretization schemes, and confirms that this method generates no spurious eigenvalues.

\begin{figure}[h]
\centering \includegraphics[height=8.0cm]{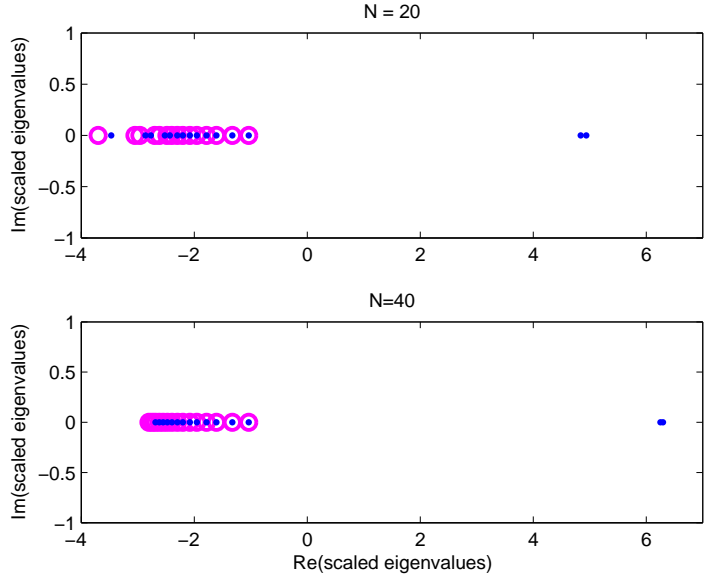}
\caption{\label{orz_eig} Eigenvalues, scaled by $x \rightarrow sign(x) + \log_{10}(1 + |x|)$, of the Gottlieb-Orszag model for $N = 20$(a) and $N = 40$ (b) discretization points. Open circles are computed using a descriptor Chebyshev Tau basis function scheme, while closed circles are computed using a traditional Chebyshev Tau basis function scheme.  Two large, positive spurious eigenvalues occur in the traditional scheme and increase with increasing $N$. They are eliminated in the descriptor scheme.}
\end{figure}

Figure~\ref{orz_eig} shows the eigenvalues of the Gottlieb-Orszag model for 20 discretization points (Fig.~\ref{orz_eig}(a)) and 40 discretization points (Fig.~\ref{orz_eig}(b)), and confirms that the descriptor method eliminates spurious eigenvalues. Open circles, computed using a descriptor method, are all in the left-hand plane. In contrast, the points computed using the general Chebyshev tau method (closed circles) include large spurious eigenvalues that increase with increasing $N$.

\begin{table}
\begin{center}
\label{compare_descriptor}
\begin{tabular}{|l|l|l|l|}\hline
{\bf Numerical Method} & $\boldsymbol{\lambda_1} $ & $\boldsymbol{\lambda_2}$ & $\boldsymbol{\lambda_3}$ \\ \hline
Exact & -9.8696 & -20.1907 & -39.4784 \\ \hline
Descriptor Chebyshev Tau ({\em collocation})   & -9.8690 & -20.1883 & -39.4694 \\ \hline
Descriptor Chebyshev Tau ({\em basis fn}) &  -9.8696 & -20.1907 & -39.4784 \\ \hline
Traditional Chebyshev Tau ({\em basis fn}) & 56,119 & 48,515 & -9.8696 \\ \hline
Modified clamped Chebyshev Galerkin ({\em collocation})~\cite{Weideman} & -9.8696 & -20.1907 & -39.4784 \\ \hline 
\end{tabular}
\caption{Comparison of the first three non-infinite eigenvalues for the 1D viscous fluid model for several discretization schemes with $N$ = 20. Both descriptor schemes (collocation and basis function) eliminate spurious eigenvalues, while the traditional Chebyshev Tau method includes large spurious eigenvalues in the right-hand plane. Weideman and Reddy~\cite{Weideman} have developed a Galerkin collocation scheme that avoids spurious eigenvalues by using modified clamped boundary conditions; this method is specific to homogeneous BCs. See Appendix~\ref{spectral} for definitions of Tau vs. Galerkin schemes and collocation vs. basis function schemes.}
\end{center}
\end{table}

\section{Conclusions}
\label{discussion}

  The descriptor framework is a generalized eigenvalue method for hydrodynamic stability problems that eliminates spurious eigenvalues, as shown in Tables~\ref{Orr_table} and~\ref{compare_descriptor}.  It ensures that infinitely fast modes will retain formally infinite eigenvalues, even when those eigenvalues are computed numerically.  In incompressible fluid flows, there are $N$ discretized pressure variables which correspond to $N$ generalized eigenvalues which are explicitly infinite.  This method enforces only the no-slip BCs and applies them in an intuitive and unambiguous way.

  While other methods reduce the number of fields (from four to two in the Orr-Sommerfeld formulation), they do so at the expense of creating higher order derivative operators, which decreases resolution for a fixed number of grid points or basis functions. In addition, the descriptor formulation does not require inversion of a differential operator, is adaptable to different discretization schemes, and can be simply extended to problems where BCs are non-trivial.

 The descriptor framework and the traditional Orr-Sommerfeld method with modified clamped boundary conditions are complementary schemes. The latter is likely to be useful for problems with homogeneous boundary conditions where the number of discretization points is very large. The former is more flexible: it is useful when higher order operators are undesirable, inhomogeneous boundary conditions are required, or direct calculations for the pressure (or any Lagrange multiplier in a general DAE) are desired. Also, newer versions of the {\bf QZ} algorithm are quite efficient~\cite{Watkins,Kagstrom}, making computations with large matrices more feasible. 

 One area of research where descriptor notation is extremely promising is the study of a fluid interacting with a compliant boundary. In this system, no-slip conditions at the wall require that the fluid velocity match the wall velocity there. These complicated Dirichlet BCs can be applied directly to the second order differential operators in $A$. Furthermore, the fluid pressure at the boundary remains as an independent variable in the eigenvalue computation, which is advantageous because the pressure at the boundary influences wall motion. This topic is currently under investigation. 

 Utilizing descriptor notation is similar in spirit to several other methods for avoiding spurious eigenvalues which require the algebraic constraints to be discretized separately from the differential equations~\cite{Gottlieb,Gardner,Boyd_book}. Gottlieb's method utilizes a {\em shooting} algorithm for determining eigenvalues --  his algorithm was developed before efficient matrix methods were available for solving generalized eigenvalue problems with singular matrices. Gardner and Boyd also describe methods where the algebraic constraint is discretized.  The descriptor framework generalizes these ideas and presents a simple, systematic method for avoiding unphysical spurious modes by using new and efficient '{\bf QZ}' algorithms. Recently, Fornberg~\cite{Fornberg} has developed a fictitious point method for avoiding spurious eigenvalues that is applicable to problems with inhomogeneous boundary conditions.
 
  Although we have focused here on hydrodynamic stability problems, descriptor notation might be advantageous to any researchers who study stability of differential-algebraic equations using spectral methods. The simplicity and generalizability of the descriptor framework suggest that it is well-suited to stability analysis in many differential-algebraic systems, including but not limited to incompressible fluids.

\section*{Acknowledgments} This work was supported by the James S. McDonnell Foundation, the David and Lucile Packard Foundation, NSF grant number DMR-0606092 and AFOSR grant number FA9550-04-1-0207. M.L.M. acknowledges an NSF Graduate Research Fellowship. M.L.M. would like to thank B. Kagstrom, E. Dunham and B. Farrell for useful discussions. 

\appendix
\section{Derivation of the Orr-Sommerfeld operator using matrix multiplication}
\label{orr_som_app}
  For a channel flow between rigid walls, the nondimensionalized, linearized Navier-Stokes equations can be written schematically as:
\begin{eqnarray}
\label{RigidOrig}
\dot{{\boldsymbol v}} &=& A {\boldsymbol v} + Q p ; \\
\label{div_const_app}
{\cal D} {\boldsymbol v} &=& 0,
\end{eqnarray}
where $Q$ is the column operator $-\{ \partial_x, \partial_y, \partial_z \}'$, and ${\cal D}$ is the row operator $\{ \partial_x, \partial_y, \partial_z \}$.  A no-slip condition at the boundaries requires $\boldsymbol{v} = \boldsymbol{0}$, while there are no explicit BCs on $p$. The operator $A$ can be written as follows:
\begin{equation}
\label{A_matrix}
A = \begin{bmatrix} \frac{\Delta}{R} + U \partial_x & U^{'} & 0 \\
0 & \frac{\Delta}{R} + U \partial_x & 0 \\
0 & 0 & \frac{\Delta}{R} + U \partial_x  \end{bmatrix},
\end{equation}
where $R$ is the Reynolds number and $U$ is the mean flow.  The mean flow is in the $x$-direction, while the channel walls ensure the flow is non-periodic in the $y$-direction. To derive the OS equation, we first to rewrite the pressure in terms of the velocities. Because the operator ${\cal D}$ commutes with time derivatives, left multiplication by  ${\cal D}$ on Eq.~\ref{RigidOrig} results in a left-hand side which is identically zero. Therefore $p$ and ${\boldsymbol v}$ have the following relationship:
\begin{equation}
\label{PSubs}
{\cal D} \left( A {\boldsymbol v} \right) = -{\cal D} Q p.
\end{equation}
A three-by-three matrix is formed from the product of the scalar ${\cal D} Q  \equiv -\Delta \equiv -\left( \partial_x^2 + \partial_y^2 + \partial_z^2 \right)$ and the identity matrix ($I$).   Left multiplication by $\Delta I$ of Eq.~\ref{RigidOrig} results in the following equation:
\begin{equation}
\label{PSubs2}
\Delta I \dot{{\boldsymbol v}} =\Delta I  \left( A {\boldsymbol v} \right) + \Delta I  Q p.
\end{equation}
Because $\Delta I$ commutes with $Q$, the two may be interchanged in the last term of Eq.~\ref{PSubs2}. The pressure $p$ can then be removed from the equation using Eq.~\ref{PSubs}, resulting in:
\begin{equation}
\label{v_only}
\Delta I \dot{{\boldsymbol v}} = \Delta I \left(  A {\boldsymbol v} \right) +  (Q {\cal D}) \left(  A {\boldsymbol v}  \right),
\end{equation}
where $Q {\cal D}$ is a three-by-three matrix. Rewriting the velocity fields in terms of the wall-normal velocity $v_y$ and vorticity $w_y$  and operating on each equation by $\Delta^{-1}$, results in the following equations~\cite{Jovanovic_thesis}:
\begin{eqnarray}
\label{OrrEnd}
\begin{bmatrix} \dot{v_y} \\ \dot{w_y} \end{bmatrix} &=& {\cal A} \begin{bmatrix} v_y \\ w_y \end{bmatrix}, \\
\mbox{where} & & \nonumber \\
{\cal A} &\equiv& \begin{bmatrix} {\cal A}_{11} & 0 \\ {\cal A}_{21} & {\cal A}_{22} \end{bmatrix}  \nonumber \\
 &=& \begin{bmatrix} - \Delta^{-1} U \partial_x \Delta + \Delta^{-1} U^{\prime \prime} \partial_x  +\Delta^{-1}\Delta^{2}/ R & 0 \\ -U^{\prime} \partial_z & -U \partial_x + \Delta / R \end{bmatrix} .
\end{eqnarray}
The term ${\cal A}_{11}$ is the well-known {\em Orr-Sommerfeld} operator acting on the wall-normal velocity, while ${\cal A}_{12}$ and  ${\cal A}_{22}$ are referred to as the {\em coupling} and {\em Squire} operators, respectively. The eigenvalues of the OS operator can be studied as a function of the wavenumber in the $x$-direction, which we denote $\alpha$.

\section{Spectral methods}
\label{spectral}

  Differentiation matrices approximate differential operators acting on functions as matrices acting on vectors. Finite difference methods track function values at points in physical space, while spectral methods approximate functions as finite series of basis functions, and keep track of the series coefficients. Because there is some confusion in terminology for spectral methods, we briefly review definitions used in this paper.

 In general, spectral methods approximate functions by a truncated series of $N$ basis polynomials, $f_N(y) = \sum_{j=0}^{N} a_j \phi_j$. We will refer to formulations with operators that act directly on these polynomials as {\em basis function} schemes. Alternatively, spectral {\em collocation} utilizes the fact that there is a one-to-one correspondence between the coefficients of that series, $a_j$, and the values of the function at specially chosen, non-uniform grid points, $f_N(y_j)$. Each function $f(y)$ can be approximated by its values at these special set of points, $f_N(y_j)$, and operators are approximated as matrices that act on these points. The grid points are chosen so that the error associated with the approximation is {\em evanescent}, or ${\cal O}\left((1/N)^N\right)$, which is superior to finite difference methods. A particularly clear introduction to spectral collocation is given by Boyd~\cite{Boyd_book}.

 There are two main methods for applying BCs in spectral methods.  The first is basis recombination, where fields are expanded in a series of basis functions that independently satisfy the appropriate BCs~\cite{Boyd_book,Canuto}, which are called {\em Galerkin} schemes. BCs can also be enforced using boundary bordering, which has two variations.  The first variation enforces homogeneous Dirichlet BCs in spectral collocation schemes.  In this case, the first and last entries in the vector correspond to the boundary points in physical space, and one simply removes the first and last rows and columns of the differentiation matrix. Note that this reduces the size of the square differentiation matrix by two.  A second variation involves using $m$ rows of the matrix to enforce $m$ BCs explicitly, which is often referred to as the {\em Tau} method.  Note that both Tau and the Galerkin methods can be implemented with collocation {\em and} basis function schemes.

 One numerical scheme referred to often in this paper is spectral collocation with Chebyshev polynomials.  The $ij$ entry of the first derivative matrix, $\mbox{Cheb}^{(1)}_{ij}$, is given by~\cite{Boyd_book}:
\begin{eqnarray}
\mbox{Cheb}^{(1)}_{ij} &\equiv& \left\{ \begin{array}{ll}
 (1 + 2N^2) /6, & \quad i=j= 0; \\
-(1 +2 N^2)/6 ,& \quad i=j=N; \\
 -x_j /\left[ 2 (1-x_j^2) \right], & \quad i=j; \: 0 < j < N ;\\
 (-1)^{i+j} p_i / \left[ p_j( x_i - x_j) \right],  &\quad i \neq j, \end{array} \right. \\
\mbox{where} & & \nonumber \\
p_0 = p_N \! &=& 2, \quad \quad p_j = 1, \quad j \in (1 \ldots N-1 ).
\end{eqnarray}
The second-derivative spectral collocation approximation is given by the square of the first-derivative matrix:  $\mbox{Cheb}^{(2)} = \left(\mbox{Cheb}^{(1)}\right)^{2}$.

\end{document}